\documentstyle[12pt,amstex]{article}

\newfont{\bit}{cmbxti10 scaled 1728}

       \def\de{depth}

   \let\de=\delta

\let\ph=\varphi

\def\0{\over } \def\1{\vec }     \def\2{{1\over2}} \def\4{{1\over4}}
\def\5{\bar }  \def\6{\partial } \def\7#1{{#1}\llap{/}}
\def\8#1{{\textstyle{#1}}}       \def\9#1{{\bf {#1}}}
 \def\llp{\hbox to 0pt{\hss /\hskip1.5pt}}
\def\llo{\hbox to 0.2pt{\hss /}} \def\llq{\hbox to 0pt{\hss /\hskip0.5pt}}
\def\so{\supset\hbox to 0pt{\hss $\displaystyle -$\hskip1pt}}

\def\<{\langle } \def\>{\rangle }

\def \({\left( } \def \){\right) }

\def\bea{\begin{eqnarray}} \def\eea{\end{eqnarray}}
\def\beann{\begin{eqnarray*}} \def\eeann{\end{eqnarray*}}
\def\beq{\begin{equation}}
\def\eeq{\end{equation}}

\def\_#1{\underline{#1}}

  \let\and=\wedge

\def\|#1{\bigg|_{_{#1}}}

\def\enq{\end{equation} }

 \def\RR{ I \hspace*{-0.8ex} R }

 \def\tha{\vartheta(a -\rho) }
 \def\sqa{\sqrt{a^2-\rho^2} }
 \def\da{\delta(\rho- a) }
 
\def\er{e_\rho }
 \def\ef{e_\phi }
 \def\pt{\partial_t }
\begin{document}
\renewcommand{\thefootnote}{\fnsymbol{footnote}}
\newpage
\pagestyle{empty}
\begin{center}
{\LARGE {The Ultrarelativistic Kerr-Geometry \\
 and \\
its Energy-Momentum Tensor\\
}}
\vfill
\vfill

\vspace{2cm}
{\large
 Herbert BALASIN
 \footnote[7]{ e-mail: hbalasin @@ email.tuwien.ac.at}
}\\
{\em
 Institut f\"ur Theoretische Physik, Technische Universit\"at Wien\\
 Wiedner Hauptstra{\ss}e 8--10, A - 1040 Wien, AUSTRIA
 }\\[.5cm]
{\em and}\\[.5cm]
{\large Herbert NACHBAGAUER
\footnote[8]{e-mail: herby @@ lapphp1.in2p3.fr}
}\\
{\em{Laboratoire de Physique Th\'eorique}}
{\small E}N{\large S}{\Large L}{\large A}P{\small P}
\footnote{URA 14-36 du CNRS, associ\'ee \`a l'E.N.S. de Lyon,
et au L.A.P.P. (IN2P3-CNRS)\\
\hspace*{0.7cm} d'Annecy-le-Vieux}
\\
{\em Chemin de Bellevue, BP 110, F - 74941 Annecy-le-Vieux Cedex,
France}\\[.5cm]
\end{center}
\vfill
\begin{abstract}
The ultrarelativistic limit of the Schwarzschild and the
Kerr-geometry together with their respective energy-momentum
tensors is derived. The approach is based on tensor-distributions making
use of the underlying Kerr-Schild structure,
which remains stable under the ultrarelativistic boost.
\\

 \noindent
 PACS numbers: 9760L, 0250
\end{abstract}
\vfill

\rightline{{\small E}N{\large S}{\Large L}{\large A}P{\small P}-A-472/94}
\rightline{TUW 94 -- 09 }
\rightline{May 1994}

\newpage

\renewcommand{\thefootnote}{\arabic{footnote}}
\setcounter{footnote}{0}
\newpage
 \pagebreak
 \pagenumbering{arabic}
 \pagestyle{plain}

 \section*{\Large\bit Introduction}
 \par
In a classical work \cite{AS} Aichelburg and Sexl obtained the
ultrarelativistic limit of the Schwarzschild geometry.
However, they noticed that the straightforward limit of the metric tensor
itself does not exist even in the sense of distributions.
They cured this problem by introducing a coordinate transformation
whose singular behavior compensated that of the boost, thereby
obtaining a sensible limit. Moreover, they were able to calculate the
energy-momentum tensor, which provided a physical interpretation of
the limit geometry as being generated by a massless ''point''-particle.
Inspired by recent interest in the quantum-mechanical scattering
of ultrarelativistic particles \cite{tHooft,Verl} their work was
generalised by applying essentially the techniques described above
to the Kerr-Newman spacetime-family \cite{LoSa}. Although the formalism
produced finite results, especially the energy-momentum tensor of
the rotating case lacks a clear physical interpretation.\par
\noindent
The present work advocates a different route of attack to the problem,
which mainly relies on the theory of tensor-distributions and the Kerr-Schild
type of all the geometries under consideration. Starting from the recently
derived energy-momentum tensors of the Schwarzschild and the Kerr-spacetime
\cite{BaNa1,BaNa2}, we calculate their ultrarelativistic limits.
Solving the Einstein-equations leads to the corresponding
 ultrarelativistic geometries, which are again of Kerr-Schild type.
More precisely, they turn out to be pp-waves \cite{JEK}.\par
\noindent
We have organised our work in the following way:
In the first part we rederive the ultrarelativistic Schwarzschild-geometry,
focusing mainly on the ``regularisation-dependence``. The second part uses
the Kerr-Schild structure to obtain the energy momentum-tensor
of Schwarzschild and Kerr. Finally, we derive the ultrarelativistic
Kerr-metric by solving the Einstein equations.

 \section*{\Large \bit 1) Regularisation-dependence and the ultrarelativistic
Schwarzschild geometry  }
\par
In order to illustrate our approach, it is useful to briefly review the
method of Aichelburg and Sexl \cite{AS} for  boosting the
Schwarzschild--geometry from a slightly different point of view.
Let us start our investigation from  the Schwarzschild
geometry in Kerr--Schild form \cite{DKS},
\beq\label{SS}
 g_{ab} = \eta_{ab} + f\, k_a k_b, \label{1}
\end{equation}
with $r=\sqrt{ x^i x^i } $ and $ f(r)=2 m/r $. $\eta_{ab} $ denotes the
flat Minkowski  metric of the decomposition and $k^a=(1 , x^i /r ) $
the principal null direction (with respect to $\eta_{ab}$ and $g_{ab}$).
The existence of the flat background metric $\eta_{ab}$ provides us with
a natural notion of boosts as its associated isometries. From a more
physical point of view, the metric
$g_{ab}$ approaches  asymptotically $\eta_{ab} $, which allows us to choose
a boosted asymptotical observer rather than a static one.
Having in mind to apply a boost to (\ref{SS}) it is useful
to  rewrite $mr$ with respect to a general Lorentz-frame
associated with $\eta_{ab}$
$$
 m^2 r^2 = ( P \cdot x)^2 + m^2 x^2 , \quad P^a =m (1,0,0,0) .
$$
In order to obtain a sensible ultrarelativistic limit for $P^a$
one has to ensure that its square vanishes in the same way as the
boost velocity increases \cite{AS}, thereby turning $P^a$ into a
null vector $p^a$.
Thus the naive limit of (\ref{SS}) is given by
\beq
 g_{ab}=\eta_{ab} +
8 { \vartheta (px) \0 px } p_a p_b   \label{3} ,
\end{equation}
where $\vartheta$ denotes the step-function.
We note that the boosted geometry again is of Kerr--Schild type,
 the flat part remaining unchanged under Lorentz-transformations.
However,  the profile function
$ \vartheta (px) / px $ is not locally integrable and thus a
meaningless quantity even in the sense of distributions.
Aichelburg and Sexl circumvented this problem by
introducing an $m$-dependent coordinate-transformation together
with the boost and performing the distributional limit afterwards.
 However, this amounts to a continuation of $\vartheta(px)/px$ to the whole
of test-function space, whose general form is given by

\begin{equation}\label{cont}
 \left( \left[ { \vartheta (px) \0 px } \right]_f ,
\ph \right)  := \int\limits_{R^4} pdx\; \bar{ p}dx\; d^2\tilde x \,
 {\vartheta (px) \0 px } [ \ph (px) - \vartheta
(e^{ f(\tilde x,\bar p x )}  - px ) \ph (0) ]
 \end{equation}
where $\bar{p}x$ and $\tilde{x}$ denote the remaining conjugate null and
spacelike coordinates, i.e. $\bar{p}^2=0$, $\bar{p}p=-1$,
$p\tilde\partial=\bar{p}\tilde\partial=0$.
(\ref{cont}) coincides  with the original profile function everywhere except
for the plane $px=0$, more precisely for all test functions supported in
$\RR^4-\{px=0\}$.
It is important to note that the continuation (\ref{cont}) generally
contains an arbitrary function $f$, which may depend on the remaining
coordinates too.
The explicit $f$-dependence is most easily displayed by employing
the following identity
$$
\left[
\vartheta (px)\0  px \right]_f  =  \left[ \vartheta (px) \0 px  \right]_{f=0}
 - \de (px) f(\tilde x,\bar p x ).
 $$
Restricting $f(\bar{p}x,\tilde{x})$ to be independent of $\bar{p}x$ allows
us to perform the distributionally well-defined coordinate-transformation
\cite{tHooft}
$$ \hat x = x + 4 p \; \vartheta (px) \log px $$
leaving us with an Minkowskian geometry everywhere except for the
null plane $px=0$.
The final form of the ultrarelativistic metric is thus given by
\beq
 g_{ab}  = \eta_{ab}  - 8 \de (px) f(\tilde x ) p_a p_b \label{4}.
\end{equation}
 The result (\ref{4}) illustrates that the most general
 regularisation introduces an arbitrary function $f(\tilde x) $
depending on the coordinates
$\tilde x$ which contributes only on the plane of the shock wave. It
should be noted that the function $f$ cannot be determined in this approach,
without any additional information.
The situation is quite analogous to second quantized field theory,
where ambiguities are a necessary consequence of regularising the infinite
 result. As in quantum field theory, a physical normalisation condition has to
 be imposed on the theory in order not to loose predictibility. In the next
 chapter, we will use the recently calculated energy-momentum tensor
 \cite{BaNa1, BaNa2} as basis for a different approach.

 \section*{\Large\bit
 2) Energy-momentum Tensor of the ultrarelativistic Schwarzschild
and Kerr-geometry}
Our proposal to attack the problem will again rely on the Kerr-Schild
structure, i.e. the flat part of the metric, which allows to
define the notion to the concept of a boost.
This time however, we will focus on the energy-momentum
tensor and show that its ultrarelativistic limit exists
unambigously, which allows us to solve the Einstein-equations for
the metric directly. We will do this in the following for
Schwarzschild as well as for Kerr. As pointed out in a previous work
\cite{BaNa1,BaNa2} the decomposition enables us to calculate the
energy-momentum tensor of the whole Kerr-Newman spacetime-family.
 With respect to Kerr-Schild coordinates one finds
\begin{align}\label{emKN}
\mbox{Schwarzschild: }  T^a{}_b &= -m\delta^{(3)}(x)\,(\partial_t)^a\,(dt)_b
 \nonumber\\
\mbox{Kerr: }  T^a{}_b &= \frac{m\delta(z)}{8\pi}\left\{ \frac{2}{a}\left(
\left[ \frac{a^2\tha}{\sqa^3}\right ] -\frac{\tha}{\sqa} -
\da\right )(dt)^a(\pt)_b \right. \nonumber\\
&+((\pt)^a(\ef)_b - (\ef)^a (dt)_b)\left( 2\left [\frac{\rho\tha}{\sqa}\right ]
-\frac{\pi}{a}\da\right ) \nonumber\\
&+\frac{2}{a}\left( - \left[\frac{\rho^2\tha}{\sqa^3}\right ]
- \frac{\tha}{\sqa}
 + 2\da\right )(\ef)^a(\ef)_b \nonumber\\
&\left. -\frac{2}{a}\frac{ \tha}{\sqa}(\er)^a(\er)_b\right \},
\end{align}
where $\er$ and $\ef$ are the polar basis vectors of the spacelike 2-plane,
 which is orthogonal to the boost-plane.
Changing now to an arbitray Lorentz-frame associated with $\eta_{ab}$,
(\ref{emKN}) becomes
\begin{align*}
\mbox{Schwarzschild: } T^a{}_b &= \delta(Qx) \delta^{(2)}(\tilde{x})\;
 P^a P_b,\\
\mbox{Kerr: } T^a{}_b &=\frac{\delta(Qx)}{8\pi} \left\{\frac{2}{a}\left(
-\left[ \frac{a^2\tha}{\sqa^3}\right ] +\frac{\tha}{\sqa} +
\da\right )P^aP_b \right. \\
&+m(P^a(\ef)_b + (\ef)^a P_b)\left( 2\left [\frac{\rho\tha}{\sqa}\right ]
-\frac{\pi}{a}\da\right ) \\
&+\frac{2m^2}{a}\left( - \left[\frac{\rho^2\tha}{\sqa^3}\right ] -
 \frac{\tha}{\sqa} + 2\da\right )(\ef)^a(\ef)_b \\
&\left. -\frac{2m^2}{a}\frac{ \tha}{\sqa}(\er)^a(\er)_b\right \},
\end{align*}
where $Q^a$ denotes the spacelike vector spanning together with its timelike
counterpart $P^a$ the 2-plane of the boost.  The calculation of the
ultrarelativistic limit $m\to 0$ of the energy-momentum tensor now merely
reduces to replacing $P^a$ and $Q^a$ by their null limit $p^a$.
\begin{align*}
\mbox{Schwarzschild: }T^a{}_b &=\delta(px) \delta^{(2)}(\tilde{x})\; p^a p_b,\\
\mbox{Kerr: }T^a{}_b &=\frac{\delta(px)}{8\pi} \frac{2}{a}\left(
-\left[ \frac{a^2\tha}{\sqa^3}\right ] +\frac{\tha}{\sqa} +
\da\right )p^ap_b
\end{align*}
Note however, that in the Kerr-case we have restricted ourselves
to boosts along the symmetry axis.
The result for the Schwarzschild case coincides
with the one obtained in \cite{AS}. Taking into account that the
limit geometries are not only of Kerr-Schild type but moreover pp-waves,
the integration of the Einstein-equations reduces to solving a 2-dimensional
Poisson-equation, which in the case of Schwarzschild is readily integrated
to give the profile function
\begin{equation}\label{ASpro}
f(px,\tilde{x})=-8\delta(px)\log \rho.
\end{equation}
Let us emphasize that during the whole computation we neither encountered
any singularities, nor had to perform any kind of  ''subtraction''
procedure. This is due to the fact that the ultrarelativistic limit
of the energy-momentum tensor is a perfectly well-defined quantity.
The singularities encountered in the usual approach are mainly due
to the fact that the ultrarelativistic boost changes the boundary
conditions for the metric, which is no longer asymptotically flat.
The situation is very similar to the one encountered in calculating
the Green-function of the Laplace equation in an arbitrary number,
 say $n$, of dimensions. Throwing
away the constant part of the solution by imposing natural boundary
conditions, renders the limit $n\to 2$ singular, since the logarithm
does not vanish at infinity. However, the concept of the delta function
remains same in  any number of dimensions.

  \section*{\Large\bit 3) Ultrarelativistic Kerr-geometry}
Having now calculated the energy-momentum tensor of the ultrarelativistic
Kerr-geometry, the corresponding metric is obtained by
solving the Einstein equations. As pointed out in the previous
paragraph, due to the pp-wave character of the geometry this reduces
to a 2-dimensional Poisson-equation for the profile function $f$.
Splitting $f(px,\tilde{x})$ into $f(\tilde{x})\delta(px)$ leaves
us  with
\begin{equation}\label{Poisson}
\tilde{\Delta } f = 4\left( \left[
\frac{a\tha}{\sqa^3}\right ]
- \frac{1}{a}\da -\frac{1}{a}
\frac{\tha}{\sqa}\right ).
\end{equation}
Integration of this equation is most easily achieved by dividing
the range of $\rho$ into two regions $0< \rho < a$ and $a< \rho <\infty$,
which in distributional language amounts to a corresponding condition
on the support of the test-functions. In the first region (\ref{Poisson})
becomes
$$
\tilde{\Delta} f_1  = 4\left( \frac{a}{\sqa^3} - \frac{1}{a}
\frac{1}{\sqa} \right),
$$
which integrates to
$$
f_1 = 8 \log \left(\frac{\rho}{a + \sqa}\right ) + \frac{4}{a}\sqa
+C \log(\frac{\rho}{a}) + D,
$$
where $C,D$ denote arbitrary integration constants. In the second region
(\ref{Poisson}) is turned into the Laplace equation
$$
\tilde{\Delta} f_2 = 0 \qquad\mbox{with the general solution}
\qquad f_2 = C' \log(\frac{\rho}{a}) + D',
$$
where $a$ is included for later convenience.
Gluing together both pieces employing step functions, i.e.
$f= \tha f_1 + \vartheta (\rho -a) f_2$, yields
\begin{align*}
\tilde{\Delta}f &= 4\left[ \frac{a\tha}{\sqa^3} \right ] -\frac{4}{a}
\frac{\tha}{\sqa} + 2\pi\delta^{(2)}(\tilde{x})(8+C)\\
& +\frac{1}{a}\da(C'-C-4) +  (D'-D)\partial_i(\er^i\da ),
\end{align*}
where we made use of the identities
\begin{align*}
&\tilde{\Delta}\left(\tha\frac{1}{a}\sqa \right ) = - \frac{1}{a}
\frac{\tha}{\sqa} - \left[ \frac{a\tha}{\sqa^3}\right ] + \frac{1}{a}\da,\\
&\tilde{\Delta}\left(\tha\log\left( \frac{\rho}{a +\sqa}\right )\right ) =
\left[ \frac{a\tha}{\sqa^3} \right ] + 2\pi\delta^{(2)}(\tilde{x})
-\frac{1}{a}\da,\\
&\tilde{\Delta}\left( \tha\log\frac{\rho}{a} \right) =
2\pi\delta ^{(2)}(\tilde{x}) - \frac{1}{a}\da,\\
&\tilde{\Delta}\left( \vartheta(\rho-a)\log\frac{\rho}{a} \right) =
\frac{1}{a}\da.
\end{align*}
Imposing (\ref{Poisson}) fixes the constants to be $C=-8, D=D', C'=-8$,
which produces the ultrarelativistic profile function
\begin{align}\label{Kerrprofile}
f &= -8\log\rho +\tha\left(8\log\left( \frac{\rho}{a+\sqa}\right )
+\frac{4}{a} \sqa \right ),\nonumber\\
\end{align}
where the value of the remaining constant $D$ has been fixed to $-8\log a$
in order to produce (\ref{ASpro}) in the limit $a\to 0$.
Comparing (\ref{Kerrprofile}) with the result obtained in \cite{LoSa},
we have found a closed form of the profile function, which is a
 well-defined distribution.
This is mainly due to the fact that no singularities were
encountered during the calculation and therefore no subtraction
ambiguities arose either.
Moreover, the energy-momentum tensors of geometries, boosted and unboosted,
are in direct relation.
\section*{\Large\bit Conclusion}
In this paper we proposed a method for obtaining ultrarelativistic
versions of the Kerr and the Schwarzschild-geometry. Our approach is
mainly based on the energy-momentum tensor, whose ultrarelativistic
form is readily obtained. Finally, solving the Einstein equations
produces a well-defined ultrarelativistic geometry. \par\noindent
The main ingredient for this calculation is provided by the Kerr-Schild
decomposition of the original geometries, since it allows the
calculation of the energy-momentum tensors as well as the definition
of the boost. We believe that the main advantage of the presented
approach lies in its ''regularisation''-independence, since all
the quantities involved are well-defined distributions.
An interesting further generalisation will cover the scattering
properties of our solution and an extension to the charged
case.


\vfill
\noindent{\em Acknowledgement:\/} The authors are greatly indebted
to Prof.~P.~C.~Aichelburg for many useful discussions.

\vfill
\end{document}